\documentclass[amsmath,amssymb,superscriptaddress,twocolumn,showpacs,preprintnumbers,prl,footinbib,aps,10pt]{revtex4-1}
\usepackage{xspace}
\usepackage[T1]{fontenc}
\usepackage{textcomp}
\usepackage{txfonts}
\usepackage{bm}

\usepackage{graphicx,color,epsfig}
\usepackage{fancyhdr}

\def\bbbc{{\mathchoice {\setbox0=\hbox{$\displaystyle\rm C$}\hbox{\hbox
to0pt{\kern0.4\wd0\vrule height0.9\ht0\hss}\box0}}
{\setbox0=\hbox{$\textstyle\rm C$}\hbox{\hbox
to0pt{\kern0.4\wd0\vrule height0.9\ht0\hss}\box0}}
{\setbox0=\hbox{$\scriptstyle\rm C$}\hbox{\hbox
to0pt{\kern0.4\wd0\vrule height0.9\ht0\hss}\box0}}
{\setbox0=\hbox{$\scriptscriptstyle\rm C$}\hbox{\hbox
to0pt{\kern0.4\wd0\vrule height0.9\ht0\hss}\box0}}}}

\newcommand{\ignore}[1]{}
\newcommand{\mComment}[1]{}
\newcommand{\gComment}[1]{}
\newcommand{\jComment}[1]{}
\newcommand{\rComment}[1]{}
\newcommand{\lComment}[1]{}

\renewcommand{\gComment}[1]{\textcolor{magenta}{Gerardo: #1}}
\begin{document}


\title{
  Formation of Magnetic Microphases in Ca$_3$Co$_2$O$_6$
}


\author{Y. Kamiya}

\author{C. D. Batista}

\affiliation{%
  Theoretical Division, Center for Nonlinear Studies, Los Alamos National Laboratory, 
  Los Alamos, New Mexico 87545, USA
}%


\date{\today}

\begin{abstract}
  We study a frustrated quantum Ising model relevant for Ca$_3$Co$_2$O$_6$ that consists of a triangular lattice of  weakly-coupled ferromagnetic (FM) chains. According to our quantum Monte Carlo (QMC) simulations, the chains become FM and form a three-sublattice ``up-up-down'' structure for $T \leq T_\text{CI}$. In contrast,  long-period spin-density-wave (SDW) {\it microphases} 
are stabilized along the chains for $T_\text{CI} < T < T_c$. Our mean field solutions reveal a quasi-continuous temperature dependence of the SDW wavelength, implying the existence of metastable states that explain the very slow dynamics observed in Ca$_3$Co$_2$O$_6$. We also discuss implications of microphases for the related multiferroic compounds Ca$_3$CoMnO$_6$ and Lu$_2$MnCoO$_6$.
\end{abstract}


\maketitle


{\it Introduction.}---%
Geometric frustration, low-dimensionality, and quantum fluctuations can lead to exotic phase transitions and  states of matter~\cite{LMM2011Introduction,Diep2004frustrated} such as
the field-induced magnetization plateaus of {{Sr}{Cu}$_2$({B}{O}$_3$)$_2$}~\cite{Kageyama1999exact,Kodama2002magnetic,Sebastian2008fractalization}, 
the spin-driven ``nematic'' transition in pnictides~\cite{Xu2008Ising,Fang2008theory,Kamiya2011Dimensional,Fernandes2012Preemptive},
and dimensional reduction in {Ba}{Cu}{Si}$_2$O$_6$~\cite{Sebastian2006dimensional,Batista2007geometric}.
{{Ca}$_3${Co}$_2$O$_6$} is another example comprising a triangular lattice of FM Ising chains coupled by weak antiferromagnetic (AFM) exchanges. 
This compound exhibits field-induced magnetization steps whose heights depend on the field sweep history and rate~\cite{Kageyama1997Field-Induced,Kageyama1997Magnetic,Maignan2000Single,Hardy2004Temperature,Moyoshi2011Incommensurate}. 
We will show that this out-of-equilibrium behavior has its roots in exotic equilibrium properties 
that can be extended to the related  multiferroic compound
{Ca$_3$Co$_{2-x}$Mn$_x$O$_6$}~\cite{Choi2008Ferroelectricity, Lancaster2009Spin, Jo2009magnetization, Kiryukhin2009Order, Flint2010Spin, Ouyang2011Short-range}.

\begin{figure}[!b]
  \vspace*{-0.2cm}
  \begin{center}
    \includegraphics[bb=0 0 661 311, width=8.6cm]{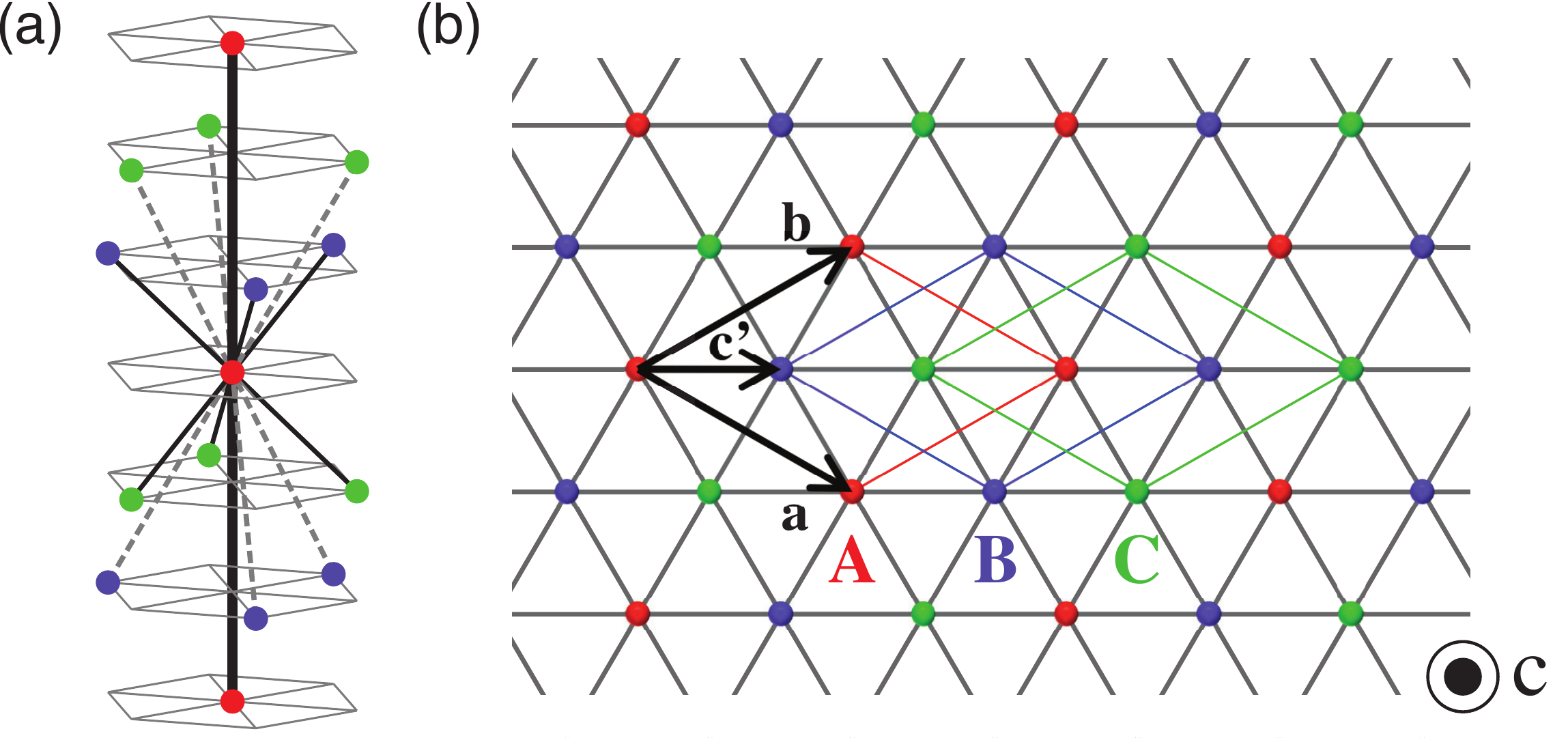}
  \end{center}
  \vspace{-0.4cm}
  \caption{%
    (Color online) 
    (a) Exchange couplings between the Co II ions:
    The FM coupling $J_1$ (a thick solid line) and the AFM couplings $J_2$ (thin solid lines) and $J_3$ (dash lines).
    The lines within layers are projections of the inter-chain couplings.
    (b) The lattice projected on the $ab$ plane. Each dot represents a chain.
  }
  \label{fig:lattice}
  \vspace{-0.1cm}
\end{figure}

The Co$^{3+}$ ions (Co II) on the trigonal prism sites of {{Ca}$_3${Co}$_2$O$_6$}
contain $3d^6$ localized electrons 
that generate an $S=2$ spin with large Ising-like anisotropy~\cite{Aasland1997Magnetic,Sampathkumaran2004Magnetic,Takubo2005Electronic,Burnus2006Valence}.
These ions  form a triangular lattice of FM Ising chains along the $c$-axis (Fig.~\ref{fig:lattice}) and 
the structure comprises  three sublattices of layers stacked along  the $c$-axis in an ${A}{B}{C}{A}{B}{C}\dots$ configuration.
Although the AFM inter-chain couplings $J_2$ and $J_3$~\cite{Fresard2004Origin} [Fig.~\ref{fig:lattice}(a)] are an order of magnitude smaller than the intra-chain FM exchange, 
$\lvert{J_1}\rvert = 2\times10\,{\rm K}$~\cite{Aasland1997Magnetic,Fresard2004Origin},
we will show that they strongly affect the intra-chain spin correlations over a window of temperatures below $T_c$.

The initial interest in {{Ca}$_3${Co}$_2$O$_6$} was triggered by the observation of 
out-of-equilibrium magnetization steps measured below $\sim 8\,{\rm K}$ and $\sim 3.6\,{\rm T}$
that appear at regular field intervals.
Previous works invoked a ``rigid-chain model'': every  chain is replaced by a single Ising spin by assuming $T \ll \lvert{J_1}\rvert$~\cite{Kudasov2006Steplike,Yao2006Monte,Kudasov2008Dynamics,Soto2009Metastable,Qin2009Two-step}.
Each spin of the resulting triangular lattice Ising model (TLIM), 
represents the magnetization of the whole chain and it is flipped if $ g \mu_B H$ overcomes its molecular field.
Within this simplified framework, the regular field intervals result from the equally-spaced discrete molecular field spectrum~\cite{Kudasov2006Steplike}.
However, this 2D scenario was challenged by the recent discovery of long-wavelength intra-chain spin-density-wave (SDW) ordering 
below $T_{c} \simeq 25\,{\rm K}$~\cite{Agrestini2008Incommensurate,Agrestini2008Nature,Moyoshi2011Incommensurate}. 
Motivated by this discovery, Chapon initiated
the study of a more realistic 3D lattice model by using a random-phase approximation (RPA) which is only valid close to $T = T_{c}$~\cite{Chapon2009Origin}.

By combining QMC simulations and mean field (MF) solutions of the 3D quantum Ising model relevant for {{Ca}$_3${Co}$_2$O$_6$}, we reproduce most of the measured zero-field properties.
A sequence of soliton lattices that lead to the observed SDW order appears for $T_{\text{CI}} < T < T_c$  through the competition between intra- and inter-chain couplings.
While the transverse field stabilizes a ferrimagnetic (FIM) up-up-down (UUD) state below $T_{\text{CI}}$ via order-by-disorder~\cite{Villain1980order,*Kamiya2009Finite}, very small longer-range exchange couplings, not included in our model, should be responsible for the actual $T=0$ ordering of {{Ca}$_3${Co}$_2$O$_6$}~\cite{Agrestini2011Slow}.
Our MF solutions show that the ordering wave-vector changes quasi-continuously as a function of $T$, implying the existence of many competing metastable states. 
Even though the modulation wavelength increases for lower $T$ and the rigid-chain picture is apparently applicable for $T<T_{\text{CI}}$, the relaxation is  known to be extremely slow and practically never complete~\cite{Moyoshi2011Incommensurate}. 
According to our results, the observed slow dynamics for $T \lesssim 10\,{\rm K}$~\cite{Moyoshi2011Incommensurate} is a direct consequence of the multiple SDW microphases that appear for $T_{\text{CI}} <  T < T_{c}$. This exotic regime can only be captured by solving the 3D model beyond the RPA~\cite{Chapon2009Origin}.

{\it Model.}---%
We use a pseudospin-1/2 to represent the lowest energy doublet ($S^{z} = \pm 2$) of the Co II ions.
The Hamiltonian is~\cite{Agrestini2008Nature,Chapon2009Origin}
\begin{align}
  {\cal H}
  &= \sum_{\left\langle{{i}{j}}\right\rangle} J_{ij} \sigma^z_i \sigma^z_{j}
  - H\sum_{i}\sigma^z_i
  - \Gamma \sum_{i} \sigma^x_i,
  \label{eq:Hamiltonian}
\end{align}
where $\bm{\sigma}_i$ is the vector of Pauli matrices for the ion $i$ on the 3D lattice shown in Fig.~\ref{fig:lattice},
$J_{{i}{j}} = J_1$ for nearest-neighbor (NN) sites along chains,
and $J_{{i}{j}} = J_2\,(J_3)$ for NN sites on NN (next-NN) layers [Fig.~\ref{fig:lattice}(a)].
Previous measurements indicate that $J_1$ is FM ($J_1<0$) while  $J_2\,,J_3 \ll \lvert{J_1}\rvert$ are AFM  \cite{Maignan2000Single}.
$H = g_c \mu_B S B$ is the external magnetic field along the $c$-axis ($g_c$ is the gyromagnetic factor), 
while the transverse field is included for accelerating the QMC relaxation 
and removing the macroscopic ground state degeneracy without invoking smaller and unknown longer-range exchange couplings.
We will assume $J_2 = J_3 = 0.1 \lvert{J_1}\rvert$, 
$\Gamma = 0.3 \lvert{J_1}\rvert$, and $H = 0$ unless otherwise specified~\footnote{
  $J_{2}/\lvert{J_1}\rvert$ and $J_{3}/\lvert{J_1}\rvert$ are reasonable choices for {Ca}$_3${Co}$_2$O$_6$ and $J_2 = J_3$ does not imply higher symmetry. 
  $\Gamma$ should be sufficiently smaller than the major exchange coupling $\lvert{J_1}\rvert$.}.

\begin{figure}
  \vspace*{-0.2cm}
  \begin{center}
    \includegraphics[bb=10 0 482 361, angle=0, width=8.6cm]{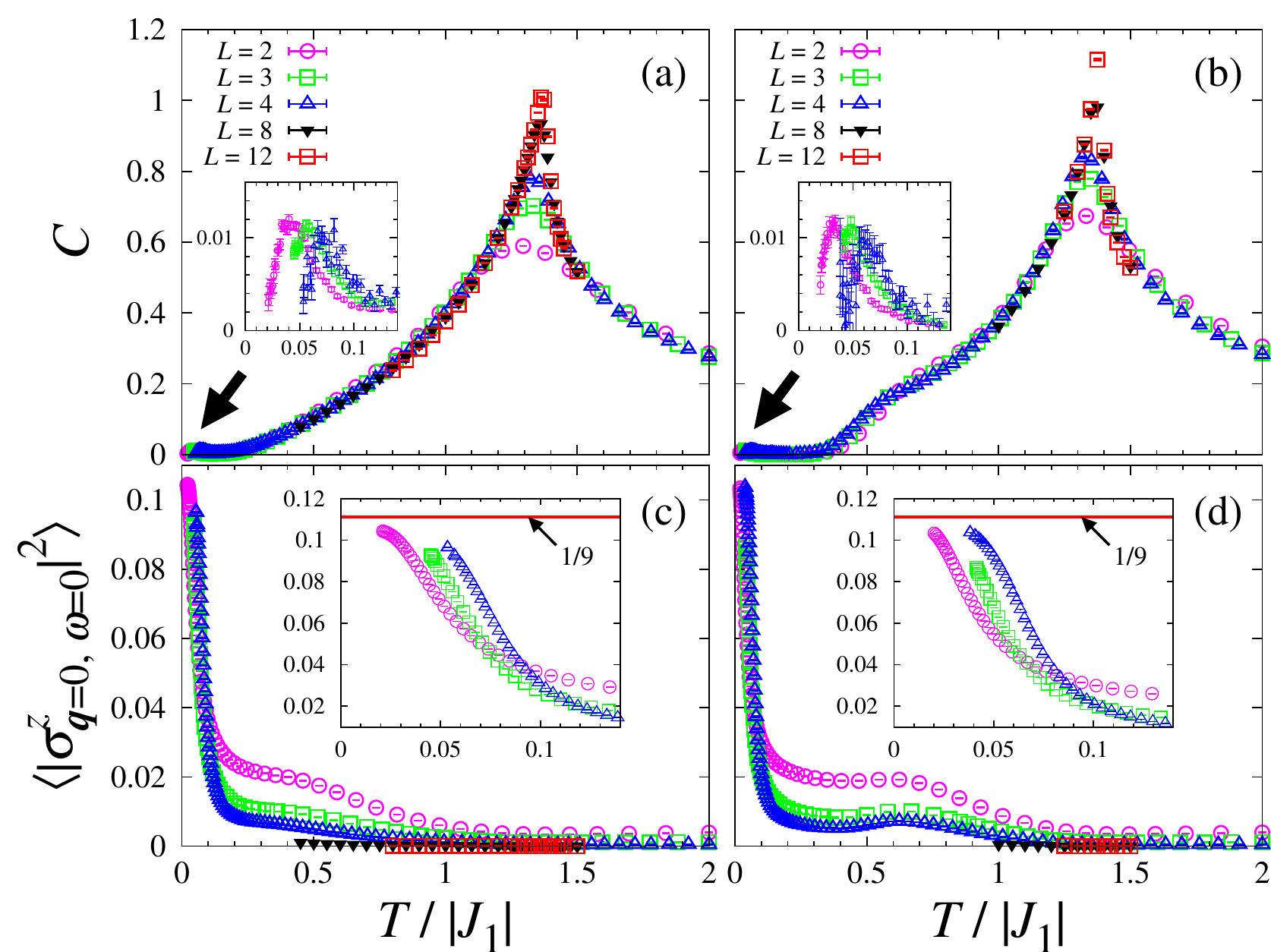}
  \end{center}
  \vspace{-0.5cm}
  \caption{%
    (Color online) 
    Upper panels show the specific heat while
    lower panels show $\lvert\sigma^z_{\bm{q}=0,\omega=0}\rvert^2$ for
    $J_2 = J_3 = 0.1 \lvert {J_1}\rvert$ and $\Gamma = 0.3 \lvert{J_1}\rvert$.
    OBC (PBC) is imposed in the $c$ direction in (a) and (c) [(b) and (d)].
    The arrows indicate a tiny anomaly in the low-$T$ regime.
    The insets provide enlarged views.
  }
  \label{fig:specific heat and magnetization}
  \vspace{-0.4cm}
\end{figure}

We use the continuous-time QMC method~\cite{Rieger1999Application,Kawashima2004Recent} to compute the thermodynamic phase diagram. 
There is no sign problem because the frustrated terms are diagonal.
The clusters can expand along each chain and the imaginary-time ($\tau$) direction. 
The weight factors due to $J_2$ and $J_3$ appear in the cluster-flip attempt. 
We use the replica exchange method~\cite{Hukushima1996Exchange} for the lowest-$T$ simulations~\footnote{
  The probability for exchanging two replicas $(T_i,\Gamma_i)$ and $(T_j,\Gamma_j)$ is 
  $p_{ij} = \min\{1,\exp(\Delta\beta\Delta{E_{\rm d}})[(\beta_j\Gamma_j)/(\beta_i\Gamma_i)]^{{\Delta}N_{\rm k}}\}$ 
  with $\Delta\beta \equiv 1/T_{i} - 1/T_{j}$.
  $\Delta{E_{\rm d}} \equiv E_{{\rm d},i} - E_{{\rm d},j}$ (${\Delta}N_{\rm k} \equiv N_{{\rm k},i} - N_{{\rm k},j}$) 
  is the difference of the diagonal term (total number of kinks in the $\tau$ direction).}.
The simulated lattice has ${L}\times{L}\times{L_c}$ unit cells with $L_c = 10L$ ( $N_{\text{layer}} \equiv 3L_c$ layers).

Figures~\ref{fig:specific heat and magnetization}(a) and \ref{fig:specific heat and magnetization}(b) show the specific heat $C$ obtained from our QMC simulations for periodic (PBC) and open boundary conditions (OBC)  along the $c$-axis, while PBC are applied in the $a$ and $b$ directions.
There are three different regimes.
The $\lambda$ peak at $T_c \simeq 1.4\lvert{J_1}\rvert$ indicates a 3D phase transition from a paramagnetic phase to an ordered state.
In addition, there are two different ordered regimes below $T_c$ separated by a tiny peak at $T \lesssim 0.1\lvert{J_1}\rvert$, which we will refer to as intermediate- and low-$T$ regimes.
While the position of this peak exhibits moderate  size dependence,
the consistent shift towards higher $T$ for larger values of $L$ implies robustness of the lowest-$T$ phase against size effects.
The sensitivity of $C(T)$ to the boundary conditions along the $c$-axis for $T \lesssim T_c$ is 
caused by a mismatch between the wave-vectors of the finite size lattice and the optimal SDW wave-vector in the intermediate-$T$ regime.
In what follows we adopt OBC along the $c$-axis because it is more convenient for detecting modulations with wavelength comparable to $N_{\text{layer}}$ (see below).

{\it FIM state in the low-$T$ regime.}---%
We will first discuss the state of equilibrium in the low-$T$ regime.
If $\Gamma = T = 0$, every chain is FM and the ground state subspace has the  well-known massive degeneracy of the TLIM~\cite{Wannier1950Antiferromagnetism}.
The lowest order correction to the ground state energy is $O(\Gamma^2)$.
We introduce the hexagonal plaquette variables $\tau_{(\mu)i}^z \equiv (1/2)\sum_{\langle{ij}\rangle_{\mu}} \sigma_j^z$ with $\langle{ij}\rangle_{\mu}$ denoting sites connected by $J_{\mu}$ ($\mu=2,3$). 
Since any unperturbed ground state satisfies $\sigma^z_i \tau^z_{(\mu)i} = -\lvert{\tau^z_{(\mu)i}}\rvert$ and $\tau_{(2)i}^z = \tau_{(3)i}^z \equiv \tau_{i}^z$, 
the energy cost of flipping a spin of the ion $i$ is  $\Delta E_i = 4\lvert{J_1}\rvert +4 \bar{J} \lvert{\tau^z_{i}}\rvert$, where $\bar{J} \equiv ({J_2 + J_3})/{2}$.
Therefore, the leading non-trivial contribution to the second-order effective Hamiltonian, $\mathcal{H}_{2}^{\text{eff}} = -\sum_i \Gamma^2/\Delta E_i$, is
\begin{align}
  \mathcal{H}_{2}^{\text{eff}} = 
  -\frac{\Gamma^2\bar{J}^2}{4\lvert{J_1}\rvert^3} 
  \sum_i  \lvert{\tau^z_{i}}\rvert^2 
  + O\left( \frac{\Gamma^2 \bar{J}^3}{\lvert{J_1}\rvert^4} \right)
  + \textrm{const.}
  \label{eq:quantum-effect}
\end{align}
Here we have used that the projection of $\sum_{i} |\tau^z_i| $  in the unperturbed ground state subspace is a constant. 
Consequently, the lowest-order non-trivial effective interaction is a FM coupling between the next-NN chains that 
stabilizes the three-sublattice UUD state of FM chains. This FIM state has a  spontaneous magnetization at 1/3 of the
saturation value. To verify this numerically, we calculate 
$\langle{\lvert{\sigma^z_{\bm{q}=0,\omega=0}}\rvert^2}\rangle 
= \langle{\lvert{N^{-1}\beta^{-1}\sum_{i}\int_{0}^{\beta}d\tau\sigma_i^z(\tau)}\rvert^2}\rangle $.
Figures~\ref{fig:specific heat and magnetization}(c) and (d) show that $\langle{\lvert{\sigma^z_{\bm{q}=0,\omega=0}}\rvert^2}\rangle$ approaches $(1/3)^2=1/9$ in the low-$T$ regime  in agreement with our analytical result. The corresponding ordering temperature coincides with the tiny anomaly in $C(T)$.

\begin{figure}
  \vspace*{-0.2cm}
  \begin{center}
    \includegraphics[bb=5 0 417 263, angle=0, width=8.6cm]{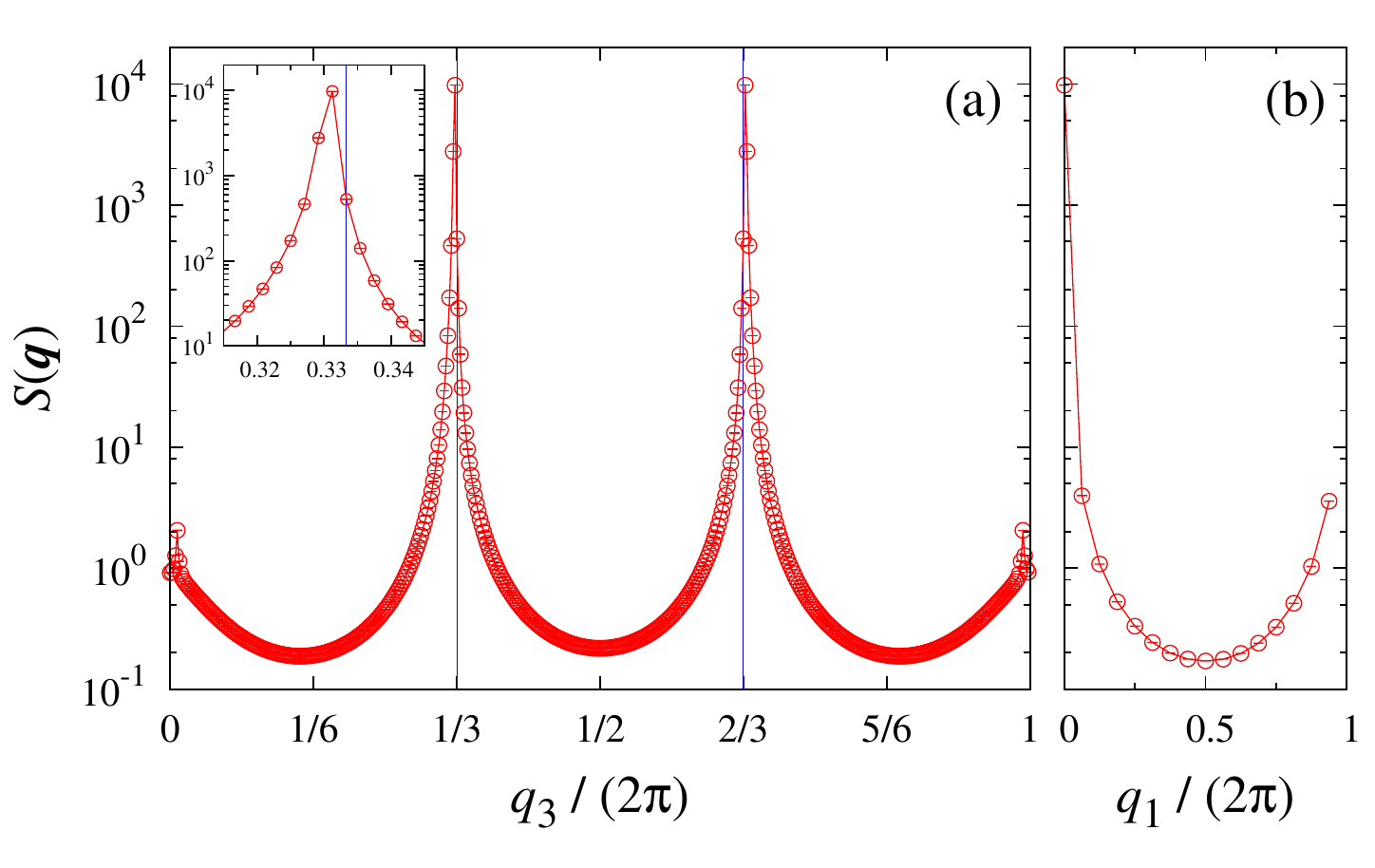}
  \end{center}
  \vspace{-0.8cm}
  \caption{%
    (Color online) 
    $S(\bm{q})$ at $T = 1.3\lvert{J_1}\rvert$ for $J_2 = J_3 = 0.1 \lvert{J_1}\rvert$, $\Gamma = 0.3 \lvert{J_1}\rvert$, $L = 16$ ($N_{\text{layer}} = 480$)  and  OBC along the $c$-axis.
    The vertical lines in (a) indicate $q_3 = 2\pi/3$ and $4\pi/3$, and the inset shows an enlarged view around $q_3 = 2\pi/3$.
    The wave-vector is varied as 
    (a) $\bm{q} = (0,0,q_3)$ and 
    (b) $\bm{q} = (q_1,0,Q_3)$.
    Error bars are smaller than the symbol size.
    The line is a guide to the eye.
  }
  \label{fig:structure-factor}
  \vspace{-0.1cm}
\end{figure}

\begin{figure}
  \vspace*{-0.2cm}
  \begin{center}
    \includegraphics[bb=5 0 360 234, angle=0, width=8.6cm]{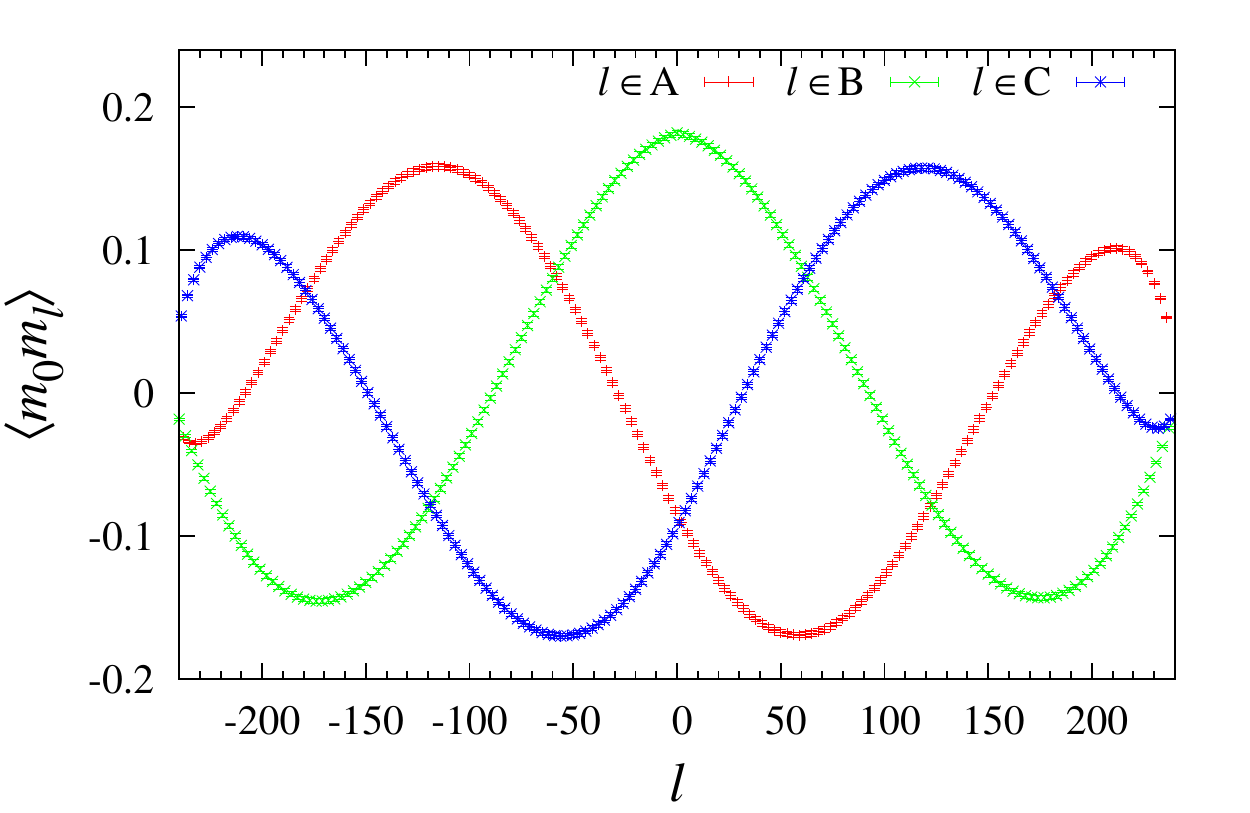}
  \end{center}
  \vspace{-0.6cm}
  \caption{%
    (Color online) 
    $\langle{m_0 m_l}\rangle$
    for the same $J_2$, $J_3$, $\Gamma$, $T$, and $L$ as in Fig.~\ref{fig:structure-factor}.
    OBC are imposed in the $c$ direction and the center layer of the simulated lattice is chosen as $l = 0$.
  }
  \label{fig:layer-moment-correlation}
  \vspace{-0.1cm}
\end{figure}

{\it SDW order in the intermediate-$T$ regime.}---%
We will next discuss the most important intermediate-$T$ regime. 
Figure~\ref{fig:structure-factor}  shows the equal-time structure factor $S(\bm{q}) = N^{-1}\sum_{ij}e^{-i\bm{q}\cdot(\bm{r}_i - \bm{r}_j)}\langle{\sigma_i^z\sigma_j^z}\rangle$ 
slightly below $T_c$ (at $T = 1.3\lvert{J_1}\rvert$), which was extracted from the major peak of $C(T)$.
The Bragg peak at $q_3 = Q_3$ is slightly shifted from $q_3 = 2\pi/3$ indicating  modulated spin ordering along the $c$-axis [Fig.~\ref{fig:structure-factor} (a)], 
while Fig.~\ref{fig:structure-factor}(b) clearly shows that each layer is FM.
This is a three-sublattice SDW order with a relative phase shift of $2\pi/3$; the numerical results are consistent with $\langle{\sigma^z_{i,\nu}}\rangle \approx a\cos(q' r_{i,3} + \phi_\nu)$ ($\nu \in \{A, B, C\}$) in the single-harmonic approximation, where $q' \equiv Q_3 - 2\pi/3$, $r_{i,3}$ is the layer index of site $i$, and $\phi_C - \phi_B = \phi_B - \phi_A = 2\pi/3$.
The very small value of $\lvert{q'}\rvert$ implies that the modulation period is very long: $2\pi\lvert{q'}\rvert^{-1} \simeq 3 \times 10^2$ at $T = 1.3\lvert{J_1}\rvert$.
These features become even more evident in the correlation functions of average moments per  layer $m_l = L^{-2}\sum_{i;\,r_{i,3} = l} \sigma_i^{z}$ (Fig.~\ref{fig:layer-moment-correlation}). 
The abrupt decay close to the edges is a consequence of OBC.
Although larger systems are necessary to determine the precise $T$-dependence of $q'$, the obtained long-period modulation is in excellent agreement with recent experiments~\cite{Agrestini2008Incommensurate,Agrestini2008Nature,Moyoshi2011Incommensurate}.


We now discuss the origin of the SDW and note that this ordering does not appear in apparently similar lattices. For example, hexagonal lattice Ising systems, such as {{Cs}{Co}{Cl}$_3$} and {{Cs}{Co}{Br}$_3$}, exhibit the partially-disordered AFM state for intermediate $T$, 
even though they are also realizations of weakly-coupled Ising FM chains that form a triangular lattice~\cite{Collins1997REVIEW}.
The crucial difference is in the connectivity of the inter-chain couplings: while the hexagonal lattice contains frustrated loops only within each layer, 
$J_2$ and $J_3$ connect spins on different layers (Fig.~\ref{fig:lattice}) and consequently compete against the dominant intra-chain coupling $J_1$.

\begin{figure}[!b]
  \vspace*{-0.3cm}
  \begin{center}
    \includegraphics[bb=5 0 360 252, angle=0, width=8.6cm]{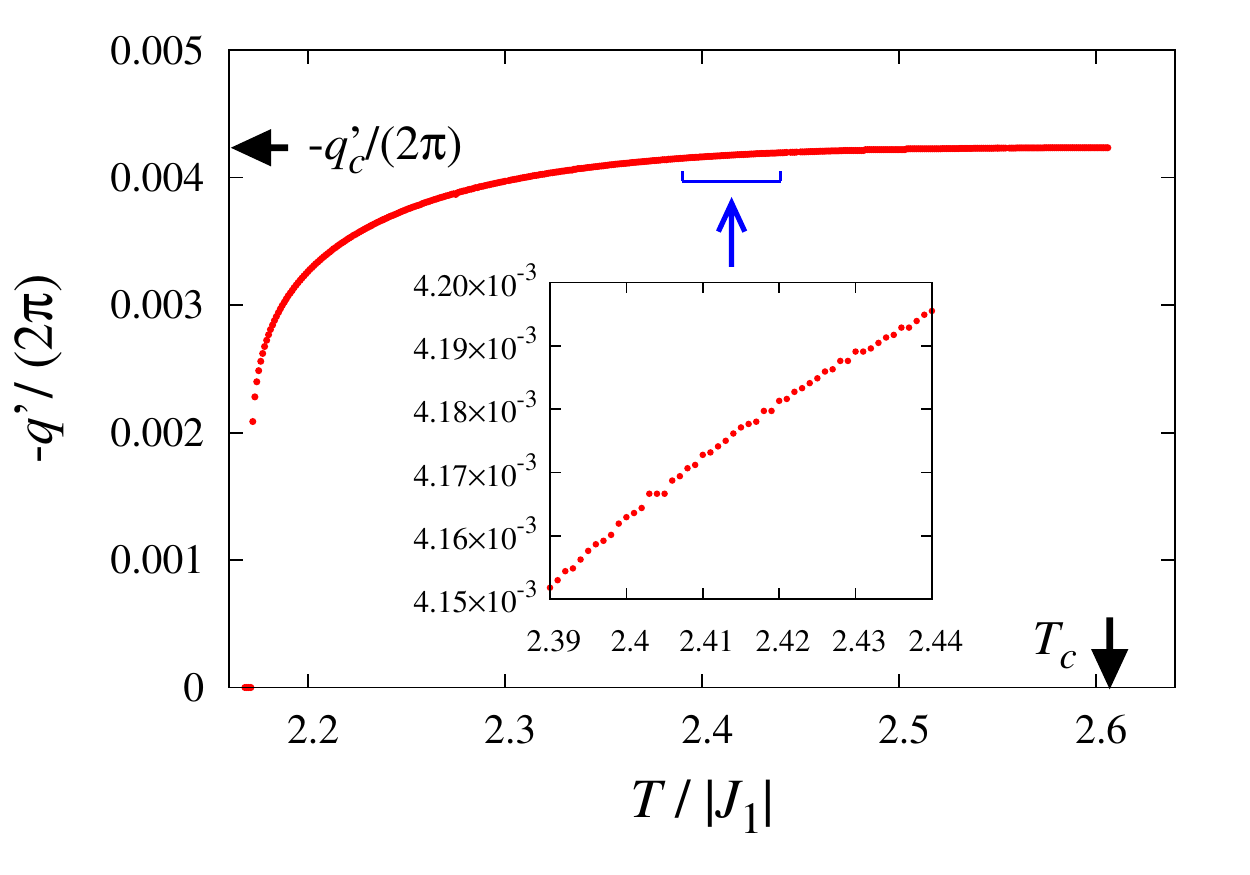}
  \end{center}
  \vspace{-0.8cm}
  \caption{%
    (Color online) 
    $q'(T)$
    obtained from the MF theory ($J_2 = J_3 = 0.1 \lvert{J_1}\rvert$).
    The inset shows an enlarged view.
  }
  \label{fig:staircase}
  \vspace{-0.1cm}
\end{figure}

Our numerical results suggest a natural MF approximation.
We assume that each layer is FM as it is indicated by our QMC simulation.
Indeed, the intra-layer effective FM  coupling is induced not only by $\Gamma$ but also by thermal fluctuations (it appears in the second-order contribution of a high-$T$ expansion). 
In fact, our MC simulations show that the microphases still exist over an extended window of temperatures even in absence of the transverse field. As expected, this phenomenon is entirely driven by the classical exchange interaction between Ising variables on a particular type of geometrically frustrated lattice.
Therefore, we will take $\Gamma=0$ in what follows.
The MF equations for the magnetization of each layer are 
\begin{align}
  \langle{m_{l}}\rangle = \tanh{\beta h_{l}} 
  \label{eq:MF}
\end{align}
with 
$h_{l} = -J_1 ( \langle{m_{l+3}}\rangle + \langle{m_{l-3}}\rangle ) - 3 J_2 ( \langle{m_{l+1}}\rangle + \langle{m_{l-1}}\rangle ) - 3 J_3 ( \langle{m_{l+2}}\rangle + \langle{m_{l-2}}\rangle ).$
The wave-vector $Q_c$ of the highest-$T$ ordered phase is given by the minimum of $J_{\text{MF}}(q) = 2J_1 \cos 3q + 6J_2 \cos q + 6J_3 \cos 2q$, and $T_c = -J_{\text{MF}}(Q_c)$~\cite{Chapon2009Origin}.
If $J_2$ and $J_3$ are AFM, any finite inter-chain coupling leads to incommensurate SDW ordering at $T = T_c$~\cite{Chapon2009Origin}.
The MF solution for $T < T_c$ is obtained by solving Eq.~\eqref{eq:MF} numerically.
Figure~\ref{fig:staircase} shows $q'(T) \equiv Q(T) - 2\pi/3$ ($Q$ is the ordering vector) obtained by imposing PBC and by varying $N_{\text{layer}}$ up to 2000 ($J_2 = J_3 = 0.1\lvert{J_1}\rvert$).
The $q'(T)$ curve is qualitatively similar to the ones obtained for the ANNNI model~\cite{Bak1982Commensurate,*Selke1988ANNNI}. However, 
we are not aware of any unambiguous realization of this prototypical model in Mott insulators.
The $q' = 0$ phase (FM chains) stabilized at the lowest temperatures is the UUD FIM state obtained from our analytical approach and from the QMC simulations. 
This state becomes unstable for $T > T_{\text{CI}} \approx 2.171\lvert{J_1}\rvert$ (the overestimation of $T_{\text{CI}}$ is expected for a MF approximation), above which $q'(T)$ changes quasi-continuously.
The obtained amplitude $\lvert{q'}\rvert$ is small in the entire regime, in agreement with experiments~\cite{Agrestini2008Incommensurate,Agrestini2008Nature,Moyoshi2011Incommensurate} and with our QMC results.
The optimal states are determined through close competition among many metastable states. 
This also implies that the fine structure of $q'(T)$ should be very sensitive to  small additional couplings that are not included in our model.
However, the quasi-continuous change of $q'(T)$ is a robust feature.

A continuum approximation of our MF theory (analogous to Ref.~\onlinecite{Bak1980Ising}) shows that the SDW phase corresponds to a quasi-continuous sequence of microphases driven
by entropic effects stabilizing a finite concentration of solitons (kinks) along the chains. 
The solitons form domain walls perpendicular to the $c$-axis.
They crystallize into a lattice and the mean separation $\lambda$ between walls 
determines $q' \propto \lambda^{-1}$. The value of $\lambda$ is controlled by a balance between the chemical potential of solitons and an effective repulsive interaction that decays exponentially in the distance between solitons~\cite{Bak1980Ising}.
The outcome of this balance is that $\lambda$ diverges logarithmically in $T - T_{\text{CI}}$ 
[our result (Fig.~\ref{fig:staircase}) reproduces this behavior].
  A number of metastable states appear in this regime with different modulation periods. 
They are separated by free energy barriers associated with the creation/annihilation and redistribution of magnetic domain walls.
These barriers give the dominant contribution to the observed slow dynamics because the relevant relaxation modes are suppressed at low $T$.

\begin{figure}
  \vspace*{-0.2cm}
  \begin{center}
    \includegraphics[bb=10 0 360 252, angle=0, width=8.6cm]{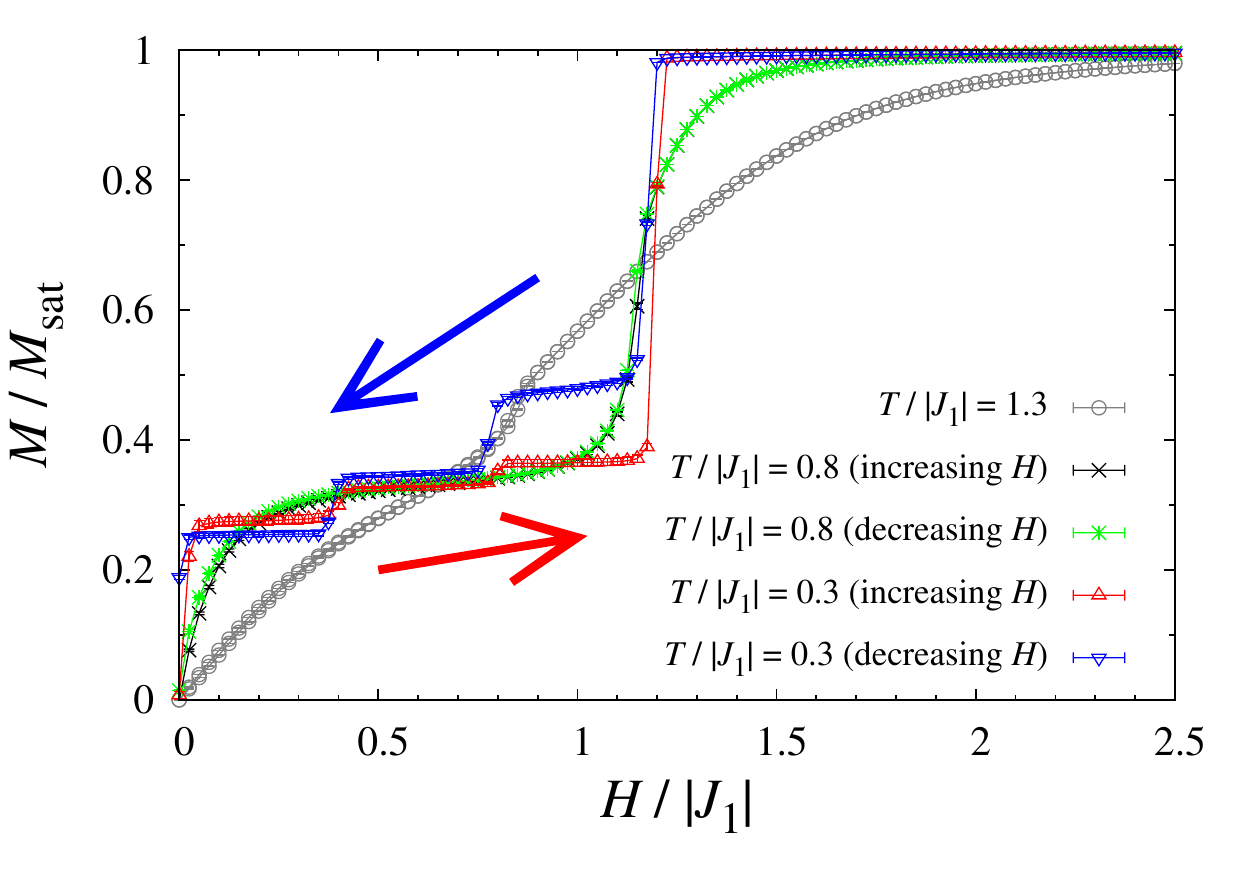}
  \end{center}
  \vspace{-0.8cm}
  \caption{%
    (Color online) 
    Magnetization curve obtained by using a relaxation process described in the text, 
    for $J_2 = J_3 = 0.1 \lvert{J_1}\rvert$, $\Gamma = 0.3 \lvert{J_1}\rvert$, $L = 12$ ($N_{\text{layer}} = 360$), and
    OBC in the $c$ direction.
  }
  \label{fig:noneq-M(H)}
  \vspace{-0.2cm}
\end{figure}

{\it Magnetization curve.}---%
Finally, we present $M(H)$ obtained in a simulated relaxation process in the realistic 3D model.
We equilibrate the system at a given $T$ for $H = 0$ and then increase $H$ gradually.
We take $10^4$ steps at each value of $H$, which is insufficient for 
equilibration at $T \ll T_c$.
After reaching a sufficiently high field, 
we go back to $H = 0$ in the same way and stop.
For $H \ne 0$, we only allow clusters to expand in the $\tau$ direction, which 
corresponds to a classical single spin flip when $\Gamma = 0$.
Although our dynamics is different from the real dynamics, our results reproduce the main experimental observations, except for the less clear steps that appear at the highest-fields (above $\sim 3.6\,{\rm T}$)~\cite{Kageyama1997Field-Induced,Kageyama1997Magnetic,Maignan2000Single,Hardy2004Temperature,Moyoshi2011Incommensurate}.
As is shown in Fig.~\eqref{fig:noneq-M(H)},
slightly below $T_c$ ($T = 1.3\lvert{J_1}\rvert$), we only find a small feature suggesting a 1/3 plateau, 
which becomes more pronounced at $T = 0.8\lvert{J_1}\rvert$ accompanied by small hysteresis.
Steps at regular magnetic field intervals appear at $T = 0.3\lvert{J_1}\rvert$, which is still inside  the SDW phase for $H = 0$. 
The reproduction of equidistant steps in the relaxation dynamics supports the notion of metastability of the observed low-$T$ states.
The heights of the steps obtained with the 3D model differ from the values obtained with the rigid-chain model~\cite{Kudasov2006Steplike,Yao2006Monte,Kudasov2008Dynamics,Soto2009Metastable,Qin2009Two-step}.

{\it Conclusions.}---%
We reproduced the temperature dependent SDW state that was reported by recent neutron diffraction experiments in {Ca}$_3${Co}$_2$O$_6$. 
More importantly, we showed that the SDW phase arises from a crystallization  of domain walls that results in a large  number of competing metastable states for $T_\text{CI} < T < T_c$. 
By uncovering these  microphases in {Ca}$_3${Co}$_2$O$_6$, we  explained the origin of the extremely slow relaxation of the Bragg peaks~\cite{Moyoshi2011Incommensurate}. Disorder induced pinning of the domain 
walls that exist in the microphases also provides a natural explanation of the observed linear-$T$ contribution in {Ca}$_3${Co}$_2$O$_6$~\cite{Hardy2003Specific}.
Order-by-disorder induced by a small transverse field leads to a FIM phase in the low-$T$ regime. However, this result does not explain the recent observation of an order-order transition to a different commensurate phase~\cite{Agrestini2011Slow}.
Therefore, although the FIM state is the ground state of Eq.~\eqref{eq:Hamiltonian}, 
other subtle perturbations, such as intra-layer AFM exchange interactions between next-NN chains, must
be included to explain the actual $T=0$ ordering of {{Ca}$_3${Co}$_2$O$_6$}.

From our results we predict that microphases should also exist in the related multiferroic compounds
{{Lu}$_2${Mn}{Co}O$_6$}~\cite{Yanez-Vilar2011Multiferroic} and {Ca$_3$Co$_{2-x}$Mn$_x$O$_6$} ($x \approx 1$)~\cite{Choi2008Ferroelectricity, Lancaster2009Spin, Jo2009magnetization, Kiryukhin2009Order, Flint2010Spin, Ouyang2011Short-range}. Since magnetic domain walls carry an internal electric dipole moment in these materials~\cite{Cheong2007Multiferroics}, the microphases should be sensitive to an external electric field that introduces a bias between walls with opposite electric polarizations.
Indeed, the dielectric constant of both compounds exhibits a broad peak below $T_c$~\cite{Jo2009magnetization,Yanez-Vilar2011Multiferroic}. We propose that this peak arises from 
the long-wavelength modulation of the electric dipole moments induced by different crystallization of magnetic domain walls (microphases).

\begin{acknowledgments}
  We thank S.-W. Cheong and T. Suzuki for valuable discussions.
  The numerical work was done in supercomputers of NERSC.
  Work at LANL was performed under the auspices of the
  U.S.\ DOE contract No.~DE-AC52-06NA25396 through the LDRD program.
\end{acknowledgments}

\bibliography{references}

\end{document}